\documentclass[a4paper,11pt]{article}
\usepackage{pos}
\usepackage{tikz}

\title{Worldline integration of photon amplitudes}
\author[a]{N. Ahmadiniaz}
\author[b]{V.M. Banda Guzmán} 
\author[c]{J.P. Edwards}
\author[a]{M.A. Lopez-Lopez}
\author[d]{C.M. Mata}
\author[e,f]{L.A. Rodriguez Chacón}
\author*[g]{C. Schubert}
\author[h]{R. Shaisultanov}

\affiliation[a]{Helmholtz-Zentrum Dresden-Rossendorf, Bautzner Landstra\ss e 400, 01328 Dresden, Germany}
\affiliation[b]{Universidad Politécnica de San Luis Potosí, Urbano Villalón 500, Colonia La Ladrillera, C.P. 78363, San Luis Potosí, San Luis Potosí, Mexico}
\affiliation[c]{Centre for Mathematical Sciences, University of Plymouth, Plymouth, PL4 8AA, UK}
\affiliation[d]{Instituto de Física y Matemáticas, Universidad Michoacana de San Nicolás de Hidalgo, Edificio C-3,Apdo. Postal 2-82, 8060 Morelia, Michoacán, Mexico}
\affiliation[e]{Computation-based Science and Technology Research Center, The Cyprus Institute,
20 Konstantinou Kavafi Street, 2121, Aglantzia, Nicosia, Cyprus}
\affiliation[f]{Department of Physics and Earth Science, University of Ferrara \& INFN, Via Saragat 1, 44122 Ferrara, Italy}
\affiliation[g]{Facultad de Ciencias Físico-Matemáticas, Universidad Michoacana de San Nicolás de Hidalgo, Avenida Francisco J. Mújica, 58060 Morelia, Michoacán, Mexico} 
\affiliation[h]{Extreme Light Infrastructure ERIC, Za Radnici 835, Dolni Brezany, 25241, Czech Republic}
\emailAdd{christianschubert137@gmail.com}

\abstract{ It has been known for many years that methods inspired by string theory, such as the worldline formalism, allow one to write down integral representations that combine large numbers of Feynman diagrams of different topologies. However, to make this fact useful for state-of-the-art calculations one has to confront non-standard integration problems where neither the known integration techniques for Feynman diagrams nor algebraic manipulation programs are of much help. Here I will give a progress report on this long-term project focussing on photon amplitudes at one and two loops, in vacuum and in external fields. }

\FullConference{Loops and Legs in Quantum Field Theory (LL2024)\\
 14-19, April, 2024\\
Wittenberg, Germany\\}

\begin{document}

\def\nonu{\nonumber\\}
\newcommand{\intT}{\int_0^\infty \dfrac{dT}{T}\, T^{4-\frac{D}{2}}\,e^{-m^2 T}\,
\int_{(4)} 
\e^{T\,\Lambda} \;}

\newcommand{\g}[1]{G_{ #1}}
\newcommand{\dg}[1]{\dot{G}_{ #1}}
\newcommand{\ddg}[1]{\ddot{G}_{ #1}}
\newcommand{\dilog}[1]{\text{Li}_2\left(#1\right)}
\newcommand{\Log}[1]{\log\left(#1\right)}
\newcommand{\hs}{\hat{s}}
\newcommand{\hT}{\hat{t}}
\newcommand{\hu}{\hat{u}}
\newcommand{\ha}{\hat{a}}
\newcommand{\hb}{\hat{b}}
\newcommand{\hc}{\hat{c}}
\newcommand{\e}{\text{e}}
\newcommand{\Arg}{\text{Arg}}
\newcommand{\bet}[1]{\beta_{\hat{#1}}}

\newcommand{\lnbeta}[1]{\ln\left( \dfrac{\beta_{#1}-1}{\beta_{#1}+1} \right)}

\newcommand{\bifmm}[3]{\bi{ -\dfrac{\beta_{#1}}{2},-\dfrac{\beta_{#2#3}}{2} }}
\newcommand{\bifmp}[3]{\bi{ -\dfrac{\beta_{#1}}{2}, \dfrac{\beta_{#2#3}}{2} }}
\newcommand{\bifpp}[3]{\bi{ \dfrac{\beta_{#1}}{2}, \dfrac{\beta_{#2#3}}{2} }}
\newcommand{\bifpm}[3]{\bi{ \dfrac{\beta_{#1}}{2}, -\dfrac{\beta_{#2#3}}{2} }}

\def\ddel{{}^\bullet\! \Delta}
\def\deld{\Delta^{\hskip -.5mm \bullet}}
\def\dddel{{}^{\bullet \bullet} \! \Delta}
\def\ddeld{{}^{\bullet}\! \Delta^{\hskip -.5mm \bullet}}
\def\deldd{\Delta^{\hskip -.5mm \bullet \bullet}}
\def\epsk#1#2{\varepsilon_{#1}\cdot k_{#2}}
\def\epseps#1#2{\varepsilon_{#1}\cdot\varepsilon_{#2}}

\newcommand{\Ascr}{\mathscr{A}}
\newcommand{\Dscr}{\mathscr{D}}
\newcommand{\Mscr}{\mathscr{M}}
\newcommand{\Wscr}{\mathscr{W}}
\newcommand{\OWscr}{\mathscr{OW}}

\newcommand{\Acal}{\mathcal{A}}
\newcommand{\Gcal}{\mathcal{G}}
\newcommand{\Dcal}{\mathcal{D}}
\newcommand{\Mcal}{\mathcal{M}}

\newcommand{\jhat}[1]{\hspace{0.3em}\widehat{\hspace{-0.4em}#1\hspace{-0.4em}}\hspace{0.4em}}
\newcommand{\bone}{1\!\!1}

%------------------------------------------------------------------------     
% MATH SYMBOLS
%
\def\cosech{\rm cosech}
\def\sech{\rm sech}
\def\coth{\rm coth}
\def\tanh{\rm tanh}
%fractions
\def\half{{1\over 2}}
\def\third{{1\over3}}
\def\fourth{{1\over4}}
\def\fifth{{1\over5}}
\def\sixth{{1\over6}}
\def\seventh{{1\over7}}
\def\eigth{{1\over8}}
\def\ninth{{1\over9}}
\def\tenth{{1\over10}}
\def\bN{\mathop{\bf N}}
\def\R{{\rm I\!R}}
\def\Eins{{\mathchoice {\rm 1\mskip-4mu l} {\rm 1\mskip-4mu l}
{\rm 1\mskip-4.5mu l} {\rm 1\mskip-5mu l}}}
\def\Z{{\mathchoice {\hbox{$\sf\textstyle Z\kern-0.4em Z$}}
{\hbox{$\sf\textstyle Z\kern-0.4em Z$}}
{\hbox{$\sf\scriptstyle Z\kern-0.3em Z$}}
{\hbox{$\sf\scriptscriptstyle Z\kern-0.2em Z$}}}}
\def\abs#1{\left| #1\right|}
\def\com#1#2{
        \left[#1, #2\right]}
\def\square{\kern1pt\vbox{\hrule height 1.2pt\hbox{\vrule width 1.2pt
   \hskip 3pt\vbox{\vskip 6pt}\hskip 3pt\vrule width 0.6pt}
   \hrule height 0.6pt}\kern1pt}
      \def\boxop{{\raise-.25ex\hbox{\square}}}
% \contract is a differential geometry contraction sign _|
\def\contract{\makebox[1.2em][c]{
        \mbox{\rule{.6em}{.01truein}\rule{.01truein}{.6em}}}}
\def\ltap{\ \raisebox{-.4ex}{\rlap{$\sim$}} \raisebox{.4ex}{$<$}\ }
\def\gtap{\ \raisebox{-.4ex}{\rlap{$\sim$}} \raisebox{.4ex}{$>$}\ }
\def\mn{{\mu\nu}}
\def\rs{{\rho\sigma}}
\newcommand{\Det}{{\rm Det}}
\def\Tr{{\rm Tr}\,}
\def\tr{{\rm tr}\,}
\def\sumij{\sum_{i<j}}
\def\e{\,{\rm e}}
%boldface vectors
\def\non{\nonumber\\}
\def\br{{\bf r}}
\def\bp{{\bf p}}
\def\bx{{\bf x}}
\def\by{{\bf y}}
\def\brhat{{\bf \hat r}}
\def\bv{{\bf v}}
\def\ba{{\bf a}}
\def\bE{{\bf E}}
\def\bB{{\bf B}}
\def\bA{{\bf A}}
%derivatives
\def\pa{\partial}
\def\dA{\partial^2}
\def\ddx{{d\over dx}}
\def\ddt{{d\over dt}}
\def\der#1#2{{d #1\over d#2}}
\def\lie{\hbox{\it \$}} % fancy L for the Lie derivative
\def\partder#1#2{{\partial #1\over\partial #2}}
\def\secder#1#2#3{{\partial^2 #1\over\partial #2 \partial #3}}
\def\kinq{{1\over 4}\dot q^2}
\def\kinb{{1\over 4}\dot x^2}
%
%equations
%\newcommand{\be}{\begin{equation}}
%\newcommand{\ee}{\end{equation}\noindent}
%\newcommand{\bear}{{\begin{eqnarray}}}
%\newcommand{\ear}{{\end{eqnarray}\noindent}}
%\newcommand{\benn}{\begin{enumerate}}
%\newcommand{\enn}{\end{enumerate}}
%\newcommand{\veject}{\vfill\eject}
%\newcommand{\ven}{\vfill\eject\noindent}
\def\bef{}
\def\ef{}
\def\be{\begin{equation}}
\def\ee{\end{equation}\noindent}
\def\bear{\begin{eqnarray}}
\def\ear{\end{eqnarray}\noindent}
\def\bec{\begin{equation}}
\def\eec{\end{equation}\noindent}
\def\bearc{\begin{eqnarray}}
\def\earc{\end{eqnarray}\noindent}
\def\benn{\begin{enumerate}}
\def\enn{\end{enumerate}}
\def\veject{\vfill\eject}
\def\ven{\vfill\eject\noindent}
%
%reference to equations
\def\eq#1{{eq. (\ref{#1})}}
\def\eqs#1#2{{eqs. (\ref{#1}) -- (\ref{#2})}}
%
%integrals
\def\totint{\int_{-\infty}^{\infty}}
\def\posint{\int_0^{\infty}}
\def\negint{\int_{-\infty}^0}
\def\pint{{\dps\int}{dp_i\over {(2\pi)}^d}}
%
% PHYS SYMBOLS
\newcommand{\GeV}{\mbox{GeV}}
\def\FFdual{F\cdot\tilde F}
\def\bra#1{\langle #1 |}
\def\ket#1{| #1 \rangle}
\def\braket#1#2{\langle {#1} \mid {#2} \rangle}
\def\vev#1{\langle #1 \rangle}
\def\rightvac{\mid 0\rangle}
\def\leftvac{\langle 0\mid}
\def\ihbar{{i\over\hbar}}
% dirac matrix stuff
\def\ge{\hbox{$\gamma_1$}}
\def\gz{\hbox{$\gamma_2$}}
\def\gd{\hbox{$\gamma_3$}}
\def\go{\hbox{$\gamma_1$}}
\def\gt{\hbox{\$\gamma_2$}}
\def\gth{\hbox{$\gamma_3$}} 
\def\gf{\hbox{$\gamma_5\;$}}
\def\slash#1{#1\!\!\!\raise.15ex\hbox {/}}
\newcommand{\slD}{\,\raise.15ex\hbox{$/$}\kern-.27em\hbox{$\!\!\!D$}}
\newcommand{\slpartial}{\raise.15ex\hbox{$/$}\kern-.57em\hbox{$\partial$}}

\newcommand{\PP}{\cal P}
\newcommand{\G}{{\cal G}}
\newcommand{\nc}{\newcommand}
\newcommand{\Fkala}{F_{\kappak_i\cdot k_j}}
\newcommand{\Fkanu}{F_{\kappa\nu}}
\newcommand{\Flaka}{F_{k_i\cdot k_j\kappa}}
\newcommand{\Flamu}{F_{k_i\cdot k_j\mu}}
\newcommand{\Fmunu}{F_{\mu\nu}}
\newcommand{\Fnumu}{F_{\nu\mu}}
\newcommand{\Fnuka}{F_{\nu\kappa}}
\newcommand{\Fmuka}{F_{\mu\kappa}}
\newcommand{\Fkalamu}{F_{\kappak_i\cdot k_j\mu}}
\newcommand{\Flamunu}{F_{k_i\cdot k_j\mu\nu}}
\newcommand{\Flanumu}{F_{k_i\cdot k_j\nu\mu}}
\newcommand{\Fkamula}{F_{\kappa\muk_i\cdot k_j}}
\newcommand{\Fkanumu}{F_{\kappa\nu\mu}}
\newcommand{\Fmulaka}{F_{\muk_i\cdot k_j\kappa}}
\newcommand{\Fmulanu}{F_{\muk_i\cdot k_j\nu}}
\newcommand{\Fmunuka}{F_{\mu\nu\kappa}}
\newcommand{\Fkalamunu}{F_{\kappak_i\cdot k_j\mu\nu}}
\newcommand{\Flakanumu}{F_{k_i\cdot k_j\kappa\nu\mu}}

\nc{\spa}[3]{\left\langle#1\,#3\right\rangle}
\nc{\spb}[3]{\left[#1\,#3\right]}
\nc{\ksl}{\not{\hbox{\kern-2.3pt $k$}}}
\nc{\hf}{\textstyle{1\over2}}
\nc{\pol}{\varepsilon}
\nc{\tq}{{\tilde q}}
\nc{\esl}{\not{\hbox{\kern-2.3pt $\pol$}}}
\newcommand{\cL}{\cal L}
\newcommand{\D}{\cal D}
\newcommand{\Dhalf}{{D\over 2}}
\def\eps{\epsilon}
\def\epshalf{{\epsilon\over 2}}
\def\lag{( -\partial^2 + V)}
%worldline
\def\freeexp{{\rm e}^{-\int_0^Td\tau {1\over 4}\dot x^2}}
\def\kinb{{1\over 4}\dot x^2}
\def\kinf{{1\over 2}\psi\dot\psi}
\def\expk{{\rm exp}\biggl[\,\sum_{i<j=1}^4 G_{Bij}k_i\cdot k_j\biggr]}
\def\expp{{\rm exp}\biggl[\,\sum_{i<j=1}^4 G_{Bij}p_i\cdot p_j\biggr]}
\def\expshort{{\e}^{\half G_{Bij}k_i\cdot k_j}}
\def\expabb{{\e}^{(\cdot )}}
\def\epseps#1#2{\varepsilon_{#1}\cdot \varepsilon_{#2}}
\def\epsk#1#2{\varepsilon_{#1}\cdot k_{#2}}
\def\kk#1#2{k_{#1}\cdot k_{#2}}
\def\G#1#2{G_{B#1#2}}
\def\Gp#1#2{{\dot G_{B#1#2}}}
\def\GF#1#2{G_{F#1#2}}
\def\Dab{{(x_a-x_b)}}
\def\Dsq{{({(x_a-x_b)}^2)}}
\def\PITD{{(4\pi T)}^{-{D\over 2}}}
\def\4piTD{{(4\pi T)}^{-{D\over 2}}}
\def\4piT4{{(4\pi T)}^{-2}}
\def\TintmD{{\dps\int_{0}^{\infty}}{dT\over T}\,e^{-m^2T}
    {(4\pi T)}^{-{D\over 2}}}
\def\Tintm4{{\dps\int_{0}^{\infty}}{dT\over T}\,e^{-m^2T}
    {(4\pi T)}^{-2}}
\def\Tintm{{\dps\int_{0}^{\infty}}{dT\over T}\,e^{-m^2T}}
\def\Tint{{\dps\int_{0}^{\infty}}{dT\over T}}
\def\np{n_{+}}
\def\nm{n_{-}}
\def\Np{N_{+}}
\def\Nm{N_{-}}
\newcommand{\slG}{{{\dot G}\!\!\!\! \raise.15ex\hbox {/}}}
\newcommand{\Gd}{{\dot G}}
\newcommand{\Gund}{{\underline{\dot G}}}
\newcommand{\Gdd}{{\ddot G}}
\def\GBd12{{\dot G}_{B12}}
\def\Dx{\dps\int{\cal D}x}
\def\Dy{\dps\int{\cal D}y}
\def\Dpsi{\dps\int{\cal D}\psi}
\def\dint#1{\int\!\!\!\!\!\int\limits_{\!\!#1}}
\def\ddtau{{d\over d\tau}}
\def\ie{\hbox{$\textstyle{\int_1}$}}
\def\iz{\hbox{$\textstyle{\int_2}$}}
\def\id{\hbox{$\textstyle{\int_3}$}}
\def\ldop{\hbox{$\lbrace\mskip -4.5mu\mid$}}
\def\rdop{\hbox{$\mid\mskip -4.3mu\rbrace$}}
%
%VARIOUS
\newcommand{\1}{{\'\i}}
\newcommand{\no}{\noindent}
\def\non{\nonumber}
\def\dps{\displaystyle}
\def\sy{\scriptscriptstyle}
\def\sy{\scriptscriptstyle}

\maketitle

\section{Worldline representation of the QED S-matrix}

\begin{figure}[htbp]
\begin{center}
 \includegraphics[width=0.31\textwidth]{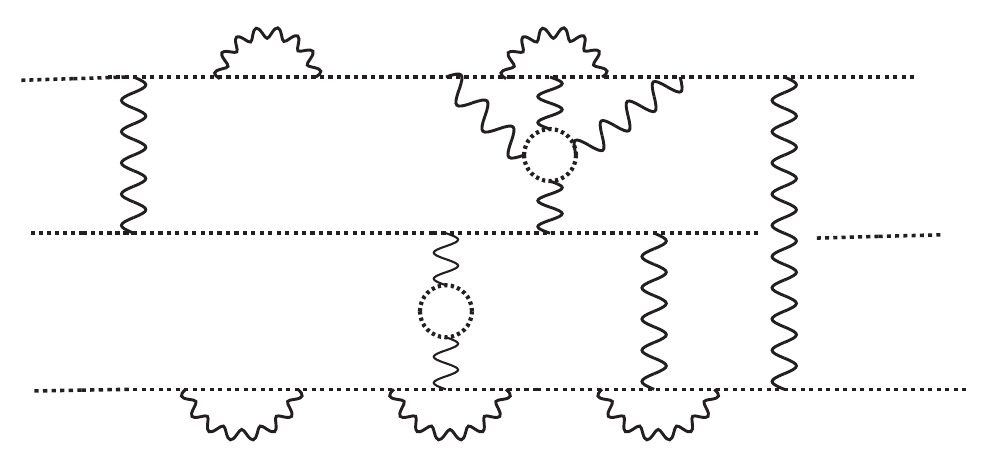}
\caption{{Contributions to the QED S-matrix with a fixed number of matter lines and loops.}}
\label{fig-QEDSmatrix}
\end{center}
\end{figure}

After Feynman's seminal work on the quantum-mechanical path integral in 1948, two years later he
used a relativistic generalization of the path integral for a perturbative construction of the S-matrices for scalar \cite{feynman1950} and spinor \cite{feynman1951} QED.
This leads to a decomposition of arbitrary QED amplitudes into contributions with a fixed number of scalar/spinor lines and loops, each one represented by
a single path integral, interconnected by photon propagators in all possible ways (Fig. \ref{fig-QEDSmatrix}). 

%Thus  the full QED S-matrix can be written in terms of first-quantized worldline path integrals.

However, after deriving the Feynman rules from this construction he seems to have concluded that they provide a much more efficient way of 
constructing QED perturbation theory than the worldline path-integral representation. This judgement was shared by the community, so that 
for several decades to come the worldline path-integral representation of QED was rarely if ever considered as a tool for everyday-life calculations of scattering amplitudes
or effective actions. It is only since the nineties, and as a spin-off of developments in string theory and supersymmetry, that it became appreciated that
this representation, although completely equivalent to the Feynman diagram one, has two principal advantages over it:

\benn

\item
It avoids the break-up of the scalar/spinor lines or loops into individual propagators.

\item
A priori it does not require a fixed ordering of the photon legs along a line or loop.

\enn

The first property becomes important when those propagators are already complicated objects by themselves, which typically happens if one wishes to
absorb an external field non-perturbatively. The second property is relevant for multi-loop calculations, since in QED the rapid growth in the number of 
diagrams with increasing number of loops is mainly due to the many different ways of inserting photons into or between fixed loops or lines. 

Let us start with the one-loop $N$-photon amplitudes in scalar QED, which in the modern ``string-inspired'' approach to Feynman's worldline formalism 
are expressed as follows \cite{polbook,strassler1,strassler2,berkosNPB} (for details, see \cite{41,126}, for generalization to the open line case \cite{dashsu}):
\bear
\Gamma[\lbrace k_i,\varepsilon_i\rbrace]
=
(-ie)^{N}
\int\! \frac{dT}{T} \e^{-m^2T}\!\!
\int \!{\cal D}x\,
V_{\rm scal}^{A}[k_1,\!\varepsilon_1]\ldots
V_{\rm scal}^{A}[k_N,\!\varepsilon_N]
\!\e^{-\int_0^Td\tau {\dot x^2\over 4}}
\, .
\label{pi}
\ear\no
Here $m$ is the mass of the loop scalar, 
and the path integral runs over the space of closed loops in (euclidean) space-time
with periodicity $T$ in proper-time. Each photon is represented by a vertex operator
\bear
V_{\rm scal}^{A}[k,\varepsilon]
=
\int_0^Td\tau\,
\varepsilon\cdot \dot x(\tau)
\,{\rm e}^{ikx(\tau)}
\, .
\ear
The zero mode  $x_0={1\over T}\int_0^Td\tau x(\tau)$  factors out of the path integral and produces the momentum conservation factor
$(2\pi)^D\delta(\sum k_i)$ which we suppress as usual. 

Since the path integral \eqref{pi} is gaussian, it can be evaluated in closed form, leading to the ``Bern-Kosower master formula''
\begin{eqnarray}
\Gamma[\lbrace k_i,\varepsilon_i\rbrace]
&=&
{(-ie)}^N
%{(2\pi )}^D\delta (\sum k_i)
{\dps\int_{0}^{\infty}}{dT\over T}
{(4\pi T)}^{-{D\over 2}}
e^{-m^2T}
\prod_{i=1}^N \int_0^T 
d\tau_i
\nonumber\\
&&\hspace{-8pt}
\times
\exp\biggl\lbrace\sum_{i,j=1}^N 
\bigl\lbrack \half G_{ij} k_i\cdot k_j
+i\dot G_{ij}k_i\cdot\varepsilon_j 
+\half\ddot G_{ij}\varepsilon_i\cdot\varepsilon_j
\bigr\rbrack\biggr\rbrace
\Big\vert_{{\rm lin}(\varepsilon_1,\ldots,\varepsilon_N)}
\label{master}
\end{eqnarray}
\no
written in terms of the ``worldline Green's function'' $G(\tau,\tau')$ and its derivatives,
\bear
G(\tau_1,\tau_2) &=& \vert \tau_1 -\tau_2\vert -{\bigl(\tau_1 -\tau_2\bigr)^2\over T}\, ,\non\\
\dot G(\tau_1,\tau_2) &=& {\rm sgn}(\tau_1 - \tau_2)
- 2 {{(\tau_1 - \tau_2)}\over T}\, ,\non\\
\ddot G(\tau_1,\tau_2)
&=& 2 {\delta}(\tau_1 - \tau_2)
- {2\over T}\, .\non
\ear
The notation $\Big\vert_{{\rm lin}(\varepsilon_1,\ldots,\varepsilon_N)}$ means projection on the terms linear in each polarization vector. 

The generalization to spinor QED can be done in various ways. Feynman's original formalism used a ``spin factor'' under the path integral
involving gamma matrices. Nowadays this spin factor is usually used only in numerical path integration, while for analytic purposes one prefers
to rewrite it as a Grassmann path integral \cite{fradkin66,fragit91}. One can then introduce worldline superfields to write down a master formula
analogous to \eqref{master}, but in practice it is usually preferable to use a certain integration-by-parts algorithm for removing the $\ddot G_{ij}$'s, 
after which the effect of spin can be implemented by the application of the ``Bern-Kosower replacement rule''

\bear
\dot G_{i_1i_2} 
\dot G_{i_2i_3} 
\cdots
\dot G_{i_ni_1}
\rightarrow 
\dot G_{i_1i_2} 
\dot G_{i_2i_3} 
\cdots
\dot G_{i_ni_1}
-
G_{Fi_1i_2}
G_{Fi_2i_3}
\cdots
G_{Fi_ni_1}\, ,
\label{reprule}
\ear
involving the { ``$\tau$-cycles''} 
$
\dot G_{i_1i_2} 
\dot G_{i_2i_3} 
\cdots
\dot G_{i_ni_1}
$
and the  fermionic worldline Green's function $G_F$, 
\bear
G_F(\tau,\tau') \equiv {\rm sgn} (\tau-\tau')
\, .
\label{defGF}
\ear 
This rule was first derived by Bern and Kosower from string theory for the case of the one-loop $N$-gluon amplitudes on-shell \cite{berkosNPB}.
Strassler \cite{strassler2} then studied the systematics of the integration-by-parts for the off-shell  case, which allowed him to see that it leads to the emergence of photon
field strength tensors $f_i^\mn = k_i^\mu \varepsilon_i^\nu - \varepsilon_i^\mu k_i^\nu$, and thus to the emergence of gauge invariance at the integrand level.
Remaining ambiguities were resolved in \cite{26} in a permutation-invariant manner. 

No such replacement rule is available for the open fermion line, for which a worldline representation suitable for state-of-the-art calculations was
found only quite recently \cite{130,131}. 

Note that the master formula \eqref{master} gives the whole amplitudes, without the need of adding ``crossed'' terms, which is made possible by the
signum and absolute-value functions appearing in the worldline Green's functions. To take full advantage of this property of the worldline formalism
it is therefore necessary to develop methods for the calculation of integrals of this type {\it without splitting them into ordered sectors}, a non-standard
integration problem for which neither existing tables of integrals nor algebraic manipulation programs are of much help. The main purpose
of my talk is to give an update on the state-of-the-art of this long-term endeavour (see also \cite{135}).

\section{Worldline representation of the four-photon amplitudes}

After the integration-by-parts, the four-photon amplitude in spinor QED appears naturally decomposed as follows:
\bear
 \Gamma_{\rm spin}(k_1,\varepsilon_1,\ldots,k_4,\varepsilon_4) 
 &=& - \frac{e^4}{8\pi^2} \Bigl( \Gamma^{(1)} +  \Gamma^{(2)}  + \Gamma^{(3)}  + \Gamma^{(4)}  +  \Gamma^{(5)} \Bigr) \,,
%\label{Gammadecomp}
\ear
\bear
\Gamma^{(1)} &=& \Gamma^{(1)}_{(1234)}T^{(1)}_{(1234)}
 + 
  \Gamma^{(1)}_{(1243)}T^{(1)}_{(1243)}
   + 
  \Gamma^{(1)}_{(1324)}T^{(1)}_{(1324)} \,,
  \nonumber\\
  \Gamma^{(2)} &=& \Gamma^{(2)}_{(12)(34)}T^{(2)}_{(12)(34)}
 + 
  \Gamma^{(2)}_{(13)(24)}T^{(2)}_{(13)(24)}
   + 
  \Gamma^{(2)}_{(14)(23)}T^{(2)}_{(14)(23)} \,,
  \nonumber\\
  \Gamma^{(3)} &=&  \sum_{i=1,2,3} \Gamma^{(3)}_{(123)i}T^{(3)r_4}_{(123)i}
  +
  \sum_{i=2,3,4} \Gamma^{(3)}_{(234)i}T^{(3)r_1}_{(234)i}
  +
  \sum_{i=3,4,1} \Gamma^{(3)}_{(341)i}T^{(3)r_2}_{(341)i}
  +
 \sum_{i=4,1,2} \Gamma^{(3)}_{(412)i}T^{(3)r_3}_{(412)i} \,,
  \nonumber\\
  \Gamma^{(4)} &=&  
\sum_{i<j}  \Gamma^{(4)}_{(ij)ii}T^{(4)}_{(ij)ii} +
\sum_{i<j}  \Gamma^{(4)}_{(ij)jj}T^{(4)}_{(ij)jj} \,,
\nonumber\\
  \Gamma^{(5)} &=&  
\sum_{i<j}  \Gamma^{(5)}_{(ij)ij}T^{(5)}_{(ij)ij} +
\sum_{i<j}  \Gamma^{(5)}_{(ij)ji}T^{(5)}_{(ij)ji} \,.
\nonumber
\ear
Here we have introduced the following tensor basis,

\bear
T^{(1)}_{(1234)} & \equiv & Z_4(1234) \, ,  \nonumber\\
T^{(2)}_{(12)(34)} & \equiv & Z_2(12)Z_2(34) \, , \nonumber\\
T^{(3)r_4}_{(123)i} & \equiv & Z_3(123) \frac{r_4\cdot f_4\cdot k_i }{r_4\cdot k_4}\, , \quad i=1,2,3 \, , \nonumber\\
T^{(4)}_{(12)ii} &\equiv & Z_2(12) \,\frac{k_i \cdot f_3 \cdot f_{4} \cdot k_i}{k_3\cdot k_4}\, , \quad i=1,2 , \nonumber\\
T^{(5)}_{(12)ij} &\equiv & Z_2(12) \, \frac{k_i \cdot f_3 \cdot f_{4} \cdot k_j}{k_3\cdot k_4}\,  ,\quad (i,j)=(1,2),(2,1)  \nonumber
\label{defTi}
\ear  
as well as
\bear
%f_i^\mn &\equiv & k_i^\mu \varepsilon_i^\nu - \varepsilon_i^\mu k_i^\nu \, , { (photon\, field\, strength \,tensor)} \nonumber\\
Z_2(ij)&\equiv&
\half {\rm tr}\bigl(f_if_j\bigr) = \varepsilon_i\cdot k_j\varepsilon_j\cdot k_i - \varepsilon_i\cdot\varepsilon_jk_i\cdot k_j \, ,
\nonumber\\
Z_n(i_1i_2\ldots i_n)&\equiv&
{\rm tr}
\Bigl(
\prod_{j=1}^n
f_{i_j}\Bigr)\, , 
\quad (n\geq 3) \, .
\quad
{ (``Lorentz\,cycle'')}
\nonumber
\label{defZn}
\ear\no
The coefficient functions are given by the following integrals,
\bear
	\Gamma^{(k)}_{\cdots}
	&=& 
	\int_0^\infty \frac{dT}{T} T^{4-\frac{D}{2}}\e^{-m^2T}
	\int_0^1\prod_{i=1}^4du_i\, \Gamma^{(k)}_{\ldots}(\Gd_{ij})\,
	\e^{\frac{1}{2}T\sum_{i,j=1}^4 G_{ij} k_i\cdot k_j}
	\nonumber
	\label{gamma}
	\ear
where we have rescaled $\tau_i = Tu_i, \, i=1,\ldots , 4$ and
\bear
\Gamma^{(1)}_{(1234)} &=& \Gd_{12}\Gd_{23}\Gd_{34}\Gd_{41} - G_{F12}G_{F23}G_{F34}G_{F41} \,, \nonumber\\
\Gamma^{(2)}_{(12)(34)} &=& \bigl(\Gd_{12}\Gd_{21} - G_{F12}G_{F21}\bigr)  \bigl(\Gd_{34}\Gd_{43} - G_{F34}G_{F43}\bigr) \,, \nonumber\\
%&&  + \bigl( \dot G_{12}\dot G_{21} - G_{F12}G_{F21}\bigr) 
%\frac{\dot G_{34}}{k_3\cdot k_4} 
%\bigl(\dot G_{41}k_4\cdot k_1 + \dot G_{42} k_4\cdot k_2 - \dot G_{31}k_3\cdot k_1 - \dot G_{32} k_3\cdot k_2 \bigr)
%\nonumber\\
%&& + 
%\frac{\dot G_{12}}{k_1\cdot k_2} 
%\bigl(\dot G_{23}k_2\cdot k_3 + \dot G_{24} k_2\cdot k_4 - \dot G_{13}k_1\cdot k_3 - \dot G_{14} k_1\cdot k_4 \bigr)
%\bigl( \dot G_{34}\dot G_{43} - G_{F34}G_{F43}\bigr)
%\nonu
\Gamma^{(3)}_{(123)1} &=& \bigl(\Gd_{12}\Gd_{23}\Gd_{31} - G_{F12}G_{F23}G_{F31}\bigr) \Gd_{41} \,, \nonumber\\
\Gamma^{(4)}_{(12)11} &=&  \bigl(\Gd_{12}\Gd_{21} - G_{F12}G_{F21}\bigr)  \Gd_{13}\Gd_{41} \,, \nonumber\\
\Gamma^{(5)}_{(12)12} &=&  \bigl(\Gd_{12}\Gd_{21} - G_{F12}G_{F21}\bigr)  \Gd_{13}\Gd_{42} \, . \nonumber\\
\label{hatgamma}
\ear
In previous work, the coefficient functions have been evaluated for the off-shell case, but with
two legs taken in the low-energy limit \cite{136,137}. In a forthcoming article \cite{inprep}, we
instead compute them for the on-shell case at full momentum, for both scalar and spinor QED.
Although this could be done using existing methods for the calculation of one-loop on-shell integrals,
with a view to eventual multi-loop generalization we prefer to do it ``the hard way'', that is,
by building tables of worldline integrals that allow one to integrate out one photon leg {\it without
fixing any ordering between the photons}. 
For starters, let us consider the basic scalar box integral. Keeping all three Mandelstam variables 
to maintain manifest permutation invariance, the universal exponent of the master formula \eqref{master} becomes
\bear
\Lambda \equiv \half \sum_{i,j=1}^4 G_{ij} k_i\cdot k_j =   -\frac{T}{2} \bigl\lbrack (G_{12} + G_{34}) s + (G_{13}+G_{24})t + (G_{14}+G_{23}) u \bigr\rbrack
\ear
Integrating out leg number 4 without fixing an ordering for the remaining three legs can be done with the formula
\bear
 \int_0^1 du_4\,\e^\Lambda &=& 
 \frac{1}{T}
\biggl\lbrack \frac{2}{u+\dot G_{12}t+\dot G_{13}s}+\frac{2}{u-\dot G_{12}t-\dot G_{13}s}\biggr\rbrack 
\,\e^{\half(G_{12}+G_{13}-G_{23})uT} 
\nonumber\\&&
+ \frac{1}{T}
\biggl\lbrack \frac{2}{t+\dot G_{23}s+\dot G_{21}u}+\frac{2}{t-\dot G_{23}s - \dot G_{21}u}\biggr\rbrack 
\,\e^{\half(G_{12}+G_{23}-G_{13})tT} 
\nonumber\\&&
+ \frac{1}{T}
\biggl\lbrack \frac{2}{s+\dot G_{31}u+\dot G_{32}t} +
\frac{2}{s-\dot G_{31}u-\dot G_{32}t} 
  \biggr\rbrack 
\,\e^{\half (G_{13}+G_{23}-G_{12})sT} 
\, .
\nonumber\\
\label{app-fullintG}
\ear
In the four-photon case, we need a generalization of this formula with additional factors of $\dot G$ in the numerator. 
For example, an additional factor of $\dot G_{41}$ leads to
\bear
 \int_0^1 du_4\, \dot{G}_{41} \e^\Lambda &=& 
\biggl\lbrack
-\frac{8}{T^2(u+\dot G_{12}t+\dot G_{13}s)^2}
 +\frac{8}{T^2(u-\dot G_{12}t-\dot G_{13}s)^2} 
 \nonumber\\&&
 +\frac{2}{T(u+\dot G_{12}t+\dot G_{13}s)}
-\frac{2}{T(u-\dot G_{12}t-\dot G_{13}s)} \biggr\rbrack 
\,\e^{\half(G_{12}+G_{13}-G_{23})uT} 
\nonumber\\&&
%+
%\biggl\lbrack
%- \frac{8}{T^2(t+\dot G_{23}s+\dot G_{21}u)^2}
%+\frac{8}{T^2(t-\dot G_{23}s - \dot G_{21}u)^2}
%\nonumber\\&&
%- \frac{2 \dot{G}_{12}}{T(t+\dot G_{23}s+\dot G_{21}u)}
%- \frac{2 \dot{G}_{12}}{T(t-\dot G_{23}s - \dot G_{21}u)}
% \biggr\rbrack 
%\,\e^{\half(G_{12}+G_{23}-G_{13})tT} 
%\nonumber\\&&
%+
%\biggl\lbrack 
%-\frac{8}{T^2(s+\dot G_{31}u+\dot G_{32}t)^2}
%+\frac{8}{T^2(s-\dot G_{31}u-\dot G_{32}t)^2} 
%\nonumber \\&&
%-\frac{2 \dot{G}_{13}}{T(s+\dot G_{31}u+\dot G_{32}t)} 
%-\frac{2 \dot{G}_{13}}{T(s-\dot G_{31}u-\dot G_{32}t)}
%  \biggr\rbrack 
%\,\e^{\half (G_{13}+G_{23}-G_{12})sT} 
%\, .
%\nonumber\\
+ 2\,{\rm Perm.}
\label{intu4one}
\ear
In \cite{inprep} along these lines we obtain the known on-shell four-photon amplitudes for spinor QED, as well as the
ones for scalar QED that apparently have never been computed in full generality. As an example, let us show the 
result for the integral of the coefficient function $\Gamma^{(5)}_{(12)12}$ in scalar QED: 
\begin{eqnarray}
\intT
\dg{12} \dg{21} \dg{13} \dg{42} 
 &=& r_{(12)12}^{(1)}  + r_{(12)12}^{(2)} \lnbeta{\hs}  + r_{(12)12}^{(3)}  \lnbeta{\hT}  \nonumber\\
&& \hspace{-200pt}
 + r_{(12)12}^{(4)}  \lnbeta{\hu} 
 + r_{(12)12}^{(5)} \left[ \lnbeta{\hs} \right]^2 + r_{(12)12}^{(6)} \left[ \lnbeta{\hT} \right]^2 + r_{(12)12}^{(7)} \left[ \lnbeta{\hu} \right]^2 \nonumber \\
&& \hspace{-200pt}
 + r_{(12)12}^{(8)} \bar{B}(s,t,u) + r_{(12)12}^{(9)} \bar{B}(s,u,t) + r_{(12)12}^{(10)} \bar{B}(t,u,s).
 \nonumber
\end{eqnarray}
Here we have introduced the variables \cite{davydychev-photonphoton}
\begin{eqnarray}
\beta_{\hat s} \equiv \sqrt{1-\frac{1}{\hat s}},\quad 
\beta_{\hat t \hat u} \equiv \sqrt{1+\frac{\hat s}{\hat t\hat u}}=\sqrt{1- \frac{1}{\hat t} - \frac{1}{\hat u}}.
\end{eqnarray}
with
$\hat s \equiv \frac{s}{4m^2},\hat t \equiv \frac{t}{4m^2},\hat u \equiv \frac{u}{4m^2}$. 
The functions  $ r_{(12)12}^{(1)},\ldots,r_{(12)12}^{(10)}$ are algebraic, and all dilogarithms are contained in a single integral $  \bar{B}(s,t,u)$, 
\begin{eqnarray}
\bar{B}(s,t,u) = \dfrac{1}{s-t} \int_0^1 dx \left[ \dfrac{s(2t+u)(1-2x)}{m^2-s(1-x)x} - \dfrac{t(2s+u)(1-2x)}{m^2-t(1-x)x}  \right] \ln\left( \dfrac{x - \dfrac{1+\beta_{\hs\hT}}{2}}{x - \dfrac{1-\beta_{\hs\hT}}{2}} \right). \label{def_Bbar}
\end{eqnarray}

\section{Incorporating a constant external field} 

The generalization of all the previous formulas from vacuum QED to the inclusion of a constant external field requires only the following changes:

\benn

\item
Replacing the worldline Green's functions $ G_B, G_F$ by field-dependent ones $ {\cal G}_B, {\cal G}_F$, 

\begin{eqnarray}
G_B(\tau_1,\tau_2) &\to &
{\cal G}_{B}(\tau_1,\tau_2) = \frac{T}{2{\cal Z}^2}
\biggl({{\cal Z}\over{{\rm sin}{\cal Z}}}
\,{\rm e}^{-i{\cal Z}\dot G_{B12}}
+i{\cal Z}\dot G_{B12} -1\biggr)\, ,
\nonumber\\
G_F(\tau_1,\tau_2) &\to &
{\cal G}_{F}(\tau_1,\tau_2) =
G_{F12}
{{\rm e}^{-i{\cal Z}\dot G_{B12}}\over {\rm cos}{\cal Z}}
\, ,
\nonumber
\end{eqnarray}
\noindent
where $ {\cal Z}_{\mu\nu} \equiv eF_{\mu\nu}T$.

\item
Adding global determinant factors
\bear
{\rm det}^{{1\over 2}}\biggl[\frac{\cal Z}{{\rm sin}{\cal Z}}\biggr] \quad {\rm (Scalar\,QED)} \, ,\quad
{\rm det}^{{1\over 2}}\biggl[\frac{\cal Z}{{\rm tan}{\cal Z}}\biggr] \quad {\rm (Spinor\,QED)} \, . \nonumber
\ear

\enn

In particular, the master formula \eqref{master} generalizes to the $N$-photon amplitudes in a constant field as 
\cite{shaisultanov,18}
\begin{eqnarray}
&&\Gamma_{\rm scal}
(k_1,\varepsilon_1;\ldots;k_N,\varepsilon_N\vert F)
=
{(-ie)}^N
%{(2\pi )}^D\delta (\sum k_i)
%\nonumber\\
%&&\hspace{-5pt}\times
{\dps\int_{0}^{\infty}}{dT\over T}
{(4\pi T)}^{-{D\over 2}}
e^{-m^2T}
{\rm det}^{{1\over 2}}
\biggl[
\frac{\cal Z}{{\rm sin}{\cal Z}}
\biggr]
\prod_{i=1}^N \int_0^T 
d\tau_i
\nonumber\\
&&\hspace{-5pt}\times
\exp\biggl\lbrace\sum_{i,j=1}^N 
\Bigl\lbrack \half k_i\cdot {\cal G}_{Bij}\cdot  k_j
-i\varepsilon_i\cdot\dot{\cal G}_{Bij}\cdot k_j
+\half
\varepsilon_i\cdot\ddot {\cal G}_{Bij}\cdot\varepsilon_j
\Bigr\rbrack\biggr\rbrace
\Big\vert_{{\rm lin}(\varepsilon_1\varepsilon_2\cdots \varepsilon_N)}\,. 
%\mid_{\rm multi-linear}\quad
\label{masterF}
\end{eqnarray}
\no
The loop replacement rule \eqref{reprule} can also be generalized in a straightforward way. 

This formalism for calculations in constant fields has already found many applications in strong-field QED. This includes
the photon propagator in a constant field \cite{ditsha,40}, magnetic photon splitting \cite{17}, 
one and two-loop Euler-Heisenberg Lagrangians \cite{5,18,51}, magnetic photon-graviton conversion \cite{61,71},
and the low-energy limit of the  $N$-photon amplitudes in a constant field \cite{156}. 

\section{Low-energy limit of $N$-photon amplitudes in vacuum and in a constant field}

Let us discuss here the last-mentioned calculation, since it is recent and relevant for our main topic, worldline integration. 
We start with the $N$-photon amplitudes in vacuum as a warm-up. 
After expanding out the exponential in the master formula \eqref{master}, and the removal of the $ \ddot G_{ij}$ 's by integration-by-parts,
the low-energy limit can simply be taken by replacing the remaining exponential factor $e^{\frac{1}{2}\sum_{i,j=1}^N G_{ij} k_i\cdot k_j}$ by unity. 
It turns out that then all terms in the integrand that are not just products of cycles turn into total derivatives. and integrate to zero.
The cycle-integrals can be done in closed form, leading to Bernoulli numbers $ B_n$:
\bear
%b_n-f_n &\equiv& 
\int_0^1 \!\!\! du_1du_2\ldots du_n\,
\Bigl(\dot G_{12}\dot G_{23}\cdots\dot G_{n1} 
-
G_{F12}G_{F23}\cdots G_{Fn1}\Bigr)
=
\left\{ \begin{array}{r@{\quad\quad}l}
2^n(2^n-2){{B}_n\over n!}  & \quad (n{\rm \quad even})\\
0 & \quad (n{\rm \quad odd})\\
\end{array} \right.
 \qquad \ 
\nonumber
\ear
This leads to the closed-form expression \cite{51}
\bear
\Gamma_{\rm spin}^{({\rm LE})}
(k_1,\varepsilon_1;\ldots ;k_N,\varepsilon_N)
&=& (-2)
\frac{e^N \Gamma(N-\frac{D}{2})}{(4\pi)^2m^{2N-D}}
\,\exp\biggl\lbrace \sum_{m=1}^{\infty}(1-2^{2m-1})\frac{b_{2m}}{2m} \tr (f_{\rm tot}^{2m})
\biggr\rbrace
\bigg \vert_{{\rm lin}(f_1\ldots f_N)}
\nonumber\\
\label{Nphotlowfinspin}
\ear
where $ f_{\rm tot} \equiv \sum_{i=1}^N f_i$, $b_n = - 2^n \frac{B_n}{n!}$. For the projection on individual helicity components, see \cite{56}. 

In the constant-field background, it is still true that the $N$-photon amplitudes can, in the weak-field limit, be reduced to terms that
factorize into cycles. However, since the generalized worldline Green's functions are non-trivial Lorentz matrices, these cycles do not any more factorize into ``$\tau$-cycles'' and ``Lorentz-cycles'', instead they
combine as

\bear
\dot G_{i_1i_2} 
\dot G_{i_2i_3} 
\cdots
\dot G_{i_ni_1}
Z_n(i_1i_2\ldots i_n)
\to
\tr \Bigl(f_{i_1}\cdot \dot {\cal G}_{Bi_1i_2} \cdot f_{i_2}
\cdot \dot {\cal G}_{Bi_2i_3} 
\cdots
f_{i_n}\cdot 
\dot {\cal G}_{Bi_ni_1}
\Bigr)
\, .
\ear
Thus the basic mathematical problem becomes the computation of the { ``open-index cycle integral''}
\bear
\int_0^1du_1\cdots \int_0^1du_n \, \dot{\cal G}_{B12}\otimes \dot{\cal G}_{B23}\otimes \cdots \otimes \dot {\cal G}_{Bn1} 
\label{oici}
\ear
which at first sight seem to generate a large number of component integrals. However, it turns out that there is a nice way of calculating them all in one go. 
Let us show this for the purely magnetic case. The magnetic worldline Green's function $\dot{\cal G}_{B}$ has the matrix decomposition \cite{40}
\bear
\dot{\cal G}_{B}(\tau_1,\tau_2)
=\dot G_{12}\,{g_-}+ S_{B12}(z)g_+ -A_{B12}(z) i{r_+}
\ear
where $ z = eBT$, $ \dot G_{12} = {\rm sgn}(\tau_1-\tau_2)$,

\bear
g_+\equiv
\left(
\begin{array}{*{4}{c}}
1&0&0&0\\
0&1&0&0\\
0&0&0&0\\
0&0&0&0
\end{array}
\right),\qquad
g_-\equiv
\left(
\begin{array}{*{4}{c}}
0&0&0&0\\
0&0&0&0\\
0&0&1&0\\
0&0&0&1
\end{array}
\right),\nonumber\\
\label{app-gd-gmat}
\ear
\vspace{-30pt}
\begin{equation}
r_+ \equiv
\left(
\begin{array}{*{4}{c}}
0&1&0&0\\
-1&0&0&0\\
0&0&0&0\\
0&0&0&0
\end{array}
\right),\qquad
r_- \equiv
\left(
\begin{array}{*{4}{c}}
0&0&0&0\\
0&0&0&0\\
0&0&0&1\\
0&0&-1&0
\end{array}
\right).
%\label{app-gd-defrmat}
%\vspace{4mm}
\nonumber
\end{equation}
and 
\bear
S_{B12}(z) &=&
{\sinh(z\,\dot G_{12})\over \sinh z} 
\, , \quad
A_{B12}(z) =
{\cosh(z \,\dot G_{12})\over 
\sinh z}-{1\over z} \;.
\nonumber
\ear
Introducing the function
\bear
H_{ij} (z) \equiv 
\frac{e^{z \dot G_{ij}}}{\sinh z} - \frac{1}{z} 
\label{defH}
\ear
the three component functions of the Green's function can be written as 
\bear
\dot G_{ij} &=& H_{ij}(0) \, , \quad
S_{Bij}(z) = \half \Bigl\lbrack H_{ij}(z) + H_{ij}(-z) \Bigr\rbrack \, ,
\quad
A_{Bij}(z) = \half \Bigl\lbrack H_{ij}(z) - H_{ij}(-z)  \Bigr\rbrack \, .
\nonumber\\
%\dot{\cal G}_{Bij}
%&=&H_{ij}(0)\,{g_-}
%+ \frac{H_{ij} (z) + H_{ij}(-z)}{2} g_
%+ -A_{B12}(z) i{r_+}
%\label{dotGbyH}
\ear
In this way the multi-component integral \eqref{oici} can be reduced to a single iterated integral,
which moreover turns out to have the following remarkable self-reproducing property:
\bear
H_{ik}^{(2)}(z,z') &\equiv &
\int_0^Td\tau_j H_{ij}(z) H_{jk}(z') = 
\frac{H_{ik}(z)}{z'-z} + \frac{H_{ik}(z')}{z-z'}
\, ,
%\label{H2}
\nonumber\\
H_{il}^{(3)}(z,z',z'') &\equiv &
\int_0^Td\tau_j \int_0^T d\tau_k H_{ij}(z) H_{jk}(z') H_{kl}(z'') 
\nonumber\\
&=&
\frac{H_{il}(z)}{(z'-z)(z''-z)}
+\frac{H_{il}(z')}{(z-z')(z''-z')}
+\frac{H_{il}(z'')}{(z-z'')(z'-z'')}
\, ,
\nonumber\\
&\vdots & \nonumber\\
H^{(n)}_{i_1i_{n+1}}(z_1,\ldots,z_n) & =&  \sum_{k=1}^n \frac{H_{i_1i_{n+1}}(z_k)}{\prod_{l \ne k} (z_l - z_k)}
\, .
%\label{H3}
\ear
Note that the right-hand sides have the full permutation symmetry.
Defining 
\bear
z_0 \equiv 0\, ,\quad z_+ \equiv z\, , \quad z_- \equiv -z\, ,
\ear
\bear
{\mathfrak g}_0 \equiv g_-\, ,\quad {\mathfrak g}_+ \equiv \half (g_+ - ir_+)\, , \quad  {\mathfrak g}_- \equiv \half (g_+ + ir_+)
\ear
we can then write
\bear
\int_0^1du_2\cdots \int_0^1du_n \, \dot{\cal G}_{B12}\otimes \dot{\cal G}_{B23}\otimes \cdots \otimes \dot {\cal G}_{Bn(n+1)} 
\!\!\!\!\!&=& \!\!\!\!\!\!\!\! \sum_{\alpha_1,\ldots,\alpha_n} H_{1(n+1)}^{(n)} (z_{\alpha_1},\ldots,z_{\alpha_n}) 
{\mathfrak g}_{\alpha_1}\otimes \cdots \otimes {\mathfrak g}_{\alpha_n}
\nonumber
\ear
where each index $\alpha_i$ runs over $0,+,-$. 
This can be extended to the spinor QED case \cite{156}, and reduces the calculation of the low-energy limit of the magnetic $ N$-photon amplitudes to { simple algebra and 
a single global proper-time integral with trigonometric integrand}.

\section{$N$-photon amplitudes in a plane-wave background}

The plane-wave background can be defined 
by a vector potential $A(x)$ of the form
\bear
e A_{\mu}(x) = a_{\mu}(n\cdot x)
\ear
where $n^{\mu}$ is a null vector, 
$ n^2 = 0$,
and as is usual we will further impose the {\it light-front gauge condition} 
$
n\cdot a = 0
$.
%Note that we absorb the charge $e$ in the definition of $a_\mu$. 
Despite of many similarities with the constant-field case, it is far from straightforward to reduce the path integrals for the $N$-photon amplitudes
to gaussian ones. Building on work of \cite{ildtor} for the two-point case, in \cite{141} the following 
master formula for the scalar QED $ N$-photon amplitude in a plane-wave background was obtained:
\bear
\Gamma_{\rm scal}(\lbrace{k_i,\varepsilon_i\rbrace};a) &=&
(-ie)^N 
(2\pi)^3 
\delta\bigl(\sum_{i=1}^N k_i^1\bigr)
\delta\bigl(\sum_{i=1}^N k_i^2\bigr)
\delta\bigl(\sum_{i=1}^N k_i^+\bigr)
\totint dx_0^+ \e^{-i x_0^+ \sum_{i=1}^N k_i^-}
\nonumber\\&&\hspace{-80pt}\times
\int_0^{\infty}
\frac{dT}{T}\,
{(4\pi T)}^{-{D\over 2}}
\prod_{i=1}^N \int_0^Td\tau_i
%\int_0^T d\tau_1 \cdots \int_0^T d\tau_N 
\e^{
\sum_{i,j=1}^N 
\bigl\lbrack  \half G_{ij} k_i\cdot k_j
-i\dot G_{ij}\varepsilon_i\cdot k_j
+\half\ddot G_{ij}\varepsilon_i\cdot\varepsilon_j
\bigr\rbrack}
\nonumber\\&&\hspace{-80pt}\times
\e^{-\bigl(m^2+ \langle\langle a^2 \rangle\rangle - \langle\langle a \rangle\rangle^2\bigr)T+  2\sum_{i=1}^N k_i \cdot \bigl(I(\tau_i)-\langle\langle I \rangle\rangle \bigr)
-2i \sum_{i=1}^N\bigl(a(\tau_i)-\langle\langle a\rangle\rangle \bigr) \cdot \varepsilon_i
}
\Bigl\vert_{\varepsilon_1\cdots \varepsilon_N}
\ear
where we have introduced light-cone coordinates and the worldloop average
\bear
\langle\langle f \rangle\rangle
\equiv
\frac{1}{T} \int_0^Td\tau f(\tau)
\, .
\ear 
Spin was incorporated in \cite{141} using the following generalization of the vacuum correlator \eqref{defGF},
\bear
\langle \psi^\mu(\tau) \psi^\nu(\tau') \rangle = \half \mathfrak G_F^\mn(\tau,\tau'), 
%\label{wickpsigen}
\nonumber
\ear
where
\bear
\mathfrak G_F^\mn(\tau,\tau') 
\equiv 
\biggl\lbrace \delta^\mn + 2i n^\mu{\cal J}^\nu(\tau,\tau') + 2i {\cal J}^\mu(\tau',\tau)n^\nu
+ 2\Bigl\lbrack {\cal J}^2(\tau,\tau')-\frac{T^2}{4} \langle\langle a'\rangle\rangle^2\Bigr\rbrack  
n^\mu n^\nu\biggr\rbrace 
G_F(\tau,\tau')
\label{Gfplane}
\nonumber\\
\ear
and we have further defined
\bear
J_\mu(\tau) &\equiv& \int_0^\tau d\tau' \Bigl( a'_\mu(\tau') - \langle\langle a'_\mu \rangle\rangle \Bigr)\, , \quad
{\cal J}_\mu(\tau,\tau') \equiv J_\mu(\tau)-J_\mu(\tau') - \frac{T}{2}\dot G (\tau,\tau')  \langle\langle a'_\mu \rangle\rangle \, .
\nonumber
\ear
See \cite{ceir} for generalization to the open-line case. 

\section{The combined constant and plane-wave field}

The arguably most complex known background field for which the Klein-Gordon and Dirac equations can be solved in closed form is the 
combination of a constant and a plane-wave field where the directions of the magnetic and of the electric field coincide with each other
and the direction of the wave propagation \cite{redmond,batfra}. Thus this background should also permit a Bern-Kosower type master formula,
and indeed it was derived in the recent \cite{154}:
\bear
\Gamma_{\rm scal}(\lbrace{k_i,\varepsilon_i\rbrace};a,F) &=&
(-ie)^N 
(2\pi)^3 
\delta\bigl(\sum_{i=1}^N k_i^1\bigr)
\delta\bigl(\sum_{i=1}^N k_i^2\bigr)
\delta\bigl(\sum_{i=1}^N k_i^+\bigr)
\totint dx_0^+ \e^{-i x_0^+ \sum_{i=1}^N k_i^-}
\nonumber\\&&\hspace{-80pt}\times
\int_0^{\infty}
\frac{dT}{T}\,
{(4\pi T)}^{-{D\over 2}}
{\rm det}^{{1\over 2}}
\biggl[
\frac{\cal Z}{{\rm sin}{\cal Z}}
\biggr]
\prod_{i=1}^N \int_0^Td\tau_i
%\int_0^T d\tau \cdots \int_0^T d\tau_N 
\,{\rm e}^{\sum_{i,j=1}^N 
\bigl\lbrack \half k_i\cdot  {\cal G}_{Bij} \cdot k_j
-i\varepsilon_i \cdot \dot {\cal G}_{Bij}\cdot k_j
+\half\varepsilon_i\cdot \ddot {\cal G}_{Bij}\cdot\varepsilon_j
\bigr\rbrack}
\nonumber\\&&\hspace{-80pt}\times
\e^{-\bigl\lbrack m^2+ \frac{1}{2} \int_0^Td\tau\int_0^Td\tau' {\tilde a}(\tau)\cdot \ddot {\cal G}_B(\tau,\tau')\cdot {\tilde a}(\tau')\bigr\rbrack T
%+2\sum_{i=1}^N k_i \cdot \bigl(I(\tau_i)-\langle\langle I \rangle\rangle \bigr) -2i \sum_{i=1}^N\bigl(a(\tau_i)-\langle\langle a\rangle\rangle \bigr) \cdot \varepsilon_i
-\sum_{i=1}^N \int_0^Td\tau 
\bigl\lbrack {\tilde a}(\tau) \cdot\dot {\cal G}_B(\tau,\tau_i) \cdot k_i + i {\tilde a}(\tau) \cdot  \ddot {\cal G}_B (\tau,\tau_i) \cdot \varepsilon_i \bigr\rbrack
}
\Bigl\vert_{\varepsilon_1\cdots \varepsilon_N}
\ear
where now
\bear
{\tilde a}_\mu(\tau)
\equiv
a_\mu 
\Bigl(x_0^+  + n\cdot \sum_{i=1}^N\bigl [-i {\cal G}_B(\tau,\tau_i)\cdot k_i + \dot {\cal G}_B(\tau,\tau_i)\cdot \varepsilon_i\bigr]\Bigr)
%\label{12-abusechanged}
\ear
and the fermionic worldline Green's function \eqref{Gfplane} has been further generalized to 
\bear
\widetilde{\mathfrak{G}}_F(\tau,\tau') &=& 
{\cal G}_F (\tau,\tau') 
+ 2i
\Bigl\lbrack
n\otimes \tilde J(\tau) \cdot {\cal G}_F(\tau,\tau')
-{\cal G}_F(\tau,\tau')\cdot n \otimes
{\tilde J} (\tau')
\Bigr\rbrack
\nonumber\\
&&. \hspace{-30pt}
+2i
\Bigl\lbrack
{\cal G}_F(\tau,\tau') \cdot {\tilde J}(\tau') \otimes n
- {\tilde J} (\tau) \otimes n\cdot {\cal G}_F (\tau,\tau')
\Bigr\rbrack
\nonumber\\
&& \hspace{-30pt}
+ 2 {\tilde J}^2(\tau) n \otimes n\cdot {\cal G}_F(\tau,\tau') 
+ 2 {\cal G}_F(\tau,\tau')\cdot n \otimes n {\tilde J}^2(\tau')
\nonumber\\
&&\hspace{-30pt}
- 4 {\tilde J}(\tau) \cdot {\cal G}_F(\tau,\tau') \cdot {\tilde J}(\tau') n\otimes n
\nonumber\\
&&\hspace{-30pt}
-\frac{iT}{z_\parallel + \lambda z_\perp}
\Bigl\lbrack
{\cal G}_F(\tau,\tau')\cdot
\Bigl( n\otimes \langle\langle{\tilde a_\lambda}'\rangle\rangle_F m_\lambda -  \langle\langle{\tilde a_{\lambda}}'\rangle\rangle_F m_\lambda \otimes n \Bigr)
\nonumber\\
&& \hspace{60pt}
-
\Bigl( n\otimes \langle\langle{\tilde a_\lambda}'\rangle\rangle_F m_\lambda -  \langle\langle{\tilde a_\lambda}'\rangle\rangle_F m_\lambda \otimes n \Bigr)
\cdot {\cal G}_F(\tau,\tau')
\Bigr\rbrack
\nonumber\\
&&\hspace{-30pt}
 + 2\frac{T}{z_\parallel + \lambda z_\perp}\Bigl\lbrack
\langle\langle{\tilde a_\lambda}'\rangle\rangle_Fm_\lambda \cdot \tilde J(\tau'){\cal G}_F(\tau,\tau') \cdot n \otimes n 
 + \langle\langle{\tilde a_\lambda}'\rangle\rangle_Fm_\lambda \cdot \tilde J(\tau) n \otimes n \cdot {\cal G}_F(\tau,\tau')\nonumber\\
&& - (\tilde J (\tau) \cdot {\cal G}_F(\tau,\tau') \cdot m_\lambda\langle\langle{\tilde a_\lambda}'\rangle\rangle_F 
+ \langle\langle{\tilde a_\lambda}'\rangle\rangle_F m_\lambda \cdot  {\cal G}_F(\tau,\tau') \cdot \tilde J(\tau') )
n\otimes n
\Bigr\rbrack
 \nonumber\\
&&\hspace{-30pt}
+ \frac{T^2}{z_{\parallel}^2-z_{\perp}^2}
\Bigl\lbrack 
\frac{ \langle\langle{\tilde a}'\rangle\rangle_F^2}{2} 
{\cal G}_F(\tau,\tau')\cdot  n\otimes n 
+ \frac{ \langle\langle{\tilde a}'\rangle\rangle_F^2}{2} 
 n\otimes n \cdot {\cal G}_F(\tau, \tau') 
 \nonumber\\&& \hspace{40pt}
- \langle\langle{\tilde a}'\rangle\rangle_F \cdot {\cal G}_F(\tau ,\tau') \cdot \langle\langle{\tilde a}'\rangle\rangle_F
n\otimes n
\Bigr\rbrack
\, .
%+ \frac{T^2}{2}\frac{\lambda z_\lambda}{z_\parallel (z_\parallel + \lambda z_\perp)^2} 
% \langle\langle{\tilde a_{-\lambda}}'\rangle\rangle_F m_{-\lambda} \cdot m_\lambda 
%  \langle\langle{\tilde a_\lambda}'\rangle\rangle_F 
%  \Bigl\lbrack {\cal G}_F(\tau,\tau') \cdot n\otimes n - n \otimes n \cdot {\cal G}_F(\tau,\tau') \Bigr\rbrack
%  \nonumber\\
%&&  - \frac{T^2}{(z_\parallel + \lambda z_\perp)^2}
% \langle\langle{\tilde a_{-\lambda}}'\rangle\rangle_F m_{-\lambda} \cdot {\cal G}_F(\tau,\tau') \cdot m_\lambda 
%  \langle\langle{\tilde a_\lambda}'\rangle\rangle_F n\otimes n
\ear
Here $ m_\pm \equiv \frac{1}{\sqrt{2}} (1,\pm i,0,0)$ and
\bear
\tilde J^{\mu}(\tau) &\equiv& \sum_{\lambda =\pm}m_{\lambda}^\mu
 \e^{-2\frac{\tau}{T} (z_{\parallel} + \lambda z_{\perp})}
 \int_0^\tau d\bar\tau
 \Bigl( {\tilde a}'_{\lambda} (\bar\tau) - \langle\langle  {\tilde a}'_{\lambda} \rangle\rangle_F \Bigr)
 \e^{2\frac{\bar\tau}{T} (z_{\parallel} + \lambda z_{\perp})}
 \nonumber
\ear
where $ z_{\perp}=eBT$, $ z_{\parallel} = ieET$, and
\bear
 \langle\langle  {\tilde a}'_{\lambda} \rangle\rangle_F &\equiv& 
% \frac{\int_0^Td\tau {\tilde a}'_{\lambda}(\tau) \e^{-2 (z_{\parallel} + \lambda z_{\perp}) \frac{T-\tau}{T}}}
% {\int_0^T d\tau \e^{-2 (z_{\parallel} + \lambda z_{\perp}) \frac{T-\tau}{T}}}
% \nonumber\\
% & =&
\frac{2 (z_{\parallel} + \lambda z_{\perp})}{1-e^{-2 (z_{\parallel} + \lambda z_{\perp})}}
 \frac{1}{T}
  \int_0^Td\tau {\tilde a}'_{\lambda}(\tau) \e^{-2 (z_{\parallel} + \lambda z_{\perp}) \frac{T-\tau}{T}}
  \, .
  \nonumber
 \ear

\section{Multi-loop photon amplitudes}

Finally, let us come to the second interesting feature of the worldline formalism mentioned in the introduction:
dealing with the amplitude as a whole becomes important when one uses the one-loop amplitudes to construct higher-loop amplitudes by sewing. 
%
%\begin{figure}[htbp]
%\begin{center}
%\includegraphics[scale=0.45]{fig-photonphoton.pdf}
%%\caption{Six permuted Feynman diagrams for photon-photon scattering.}
%\label{fig-photonphoton}
%\end{center}
%\end{figure}
%\noindent
%
E.g., from the four-photon amplitude we can construct the two-loop photon propagator,

\vspace{-100pt}
\begin{figure}[htbp]
\begin{center}
\includegraphics[scale=0.44]{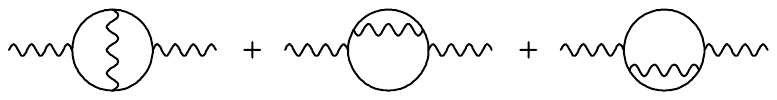}
\vspace{-240pt}
\caption{Summing diagrams for the two-loop photon propagator.}
\label{fig-2loop}
\end{center}
\end{figure}

\vspace{-20pt}

\noindent 
with all three diagrams combined into a single integral. 
Similarly, from the one-loop six-photon amplitude we get the three-loop quenched propagator

\vspace{-120pt}

\begin{figure}[htbp]
\begin{center}
\includegraphics[scale=0.34]{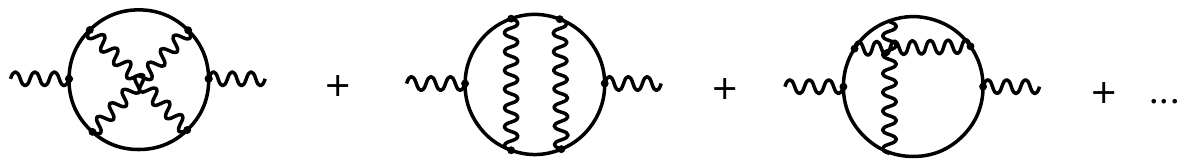}
\vspace{-130pt}
\caption{Summing diagrams for the three-loop quenched photon propagator.}
\label{fig-3loop}
\end{center}
\end{figure}
\vspace{-20pt}
\noindent
etc.  And precisely this type of sums of diagrams is known to suffer from particularly extensive cancellations (see, e.g., \cite{brdekr,135}).
More efficient than sewing is the use of  multi-loop wordline Green's functions
 that hold the information on photon insertions \cite{8,15}. For a single insertion,
%%
%\bear
%G^{(1)}_{12} &=&G_{12}-{1\over 4}
%{{\Bigl(G_{1a}-G_{1b}-G_{2a}+G_{2b}\Bigr)
%}^2
%\over
%\bar T
%+ G_{ab}}
%\, .
%\label{G2loopsub}
%\ear
%\no 
%

%
\bear
G^{(1)}(\tau_1,\tau_2)=
G(\tau_1,\tau_2) + \half
{{[G(\tau_1,\tau_a)-G(\tau_1,\tau_b)]
[G(\tau_a,\tau_2)-G(\tau_b,\tau_2)]}
\over
{{\bar T} + G(\tau_a,\tau_b)}}
\label{G2loop}
\ear
\no
where $ \bar T$ is the proper-time length of the inserted propagator, and
$ \tau_a,\tau_b$ the points on the loop between which the propagator is inserted. 
It leads to integral representations for the $ l$-loop photon propagator naturally written
in the variables 
$ G_{a_1b_1}, G_{a_2b_2},\ldots , G_{a_lb_l},C_{a_1b_1a_2b_2},\ldots , C_{a_{l-1}b_{l-1}a_lb_l}$, 
where the $ G_{a_ib_i}$ depend only on a single propagator, and the $ C_{a_ib_i a_j b_j}$ on pairs of propagators. 

In continuation of previous work \cite{15} (see also \cite{41}) on the two-loop QED beta functions, we (V.M.B.G. and C.S.) are presently calculating the
full two-loop photon propagators for both scalar and spinor QED  along the lines outlined above. Let us show here the 
integral representation obtained using the Green's function \eqref{G2loop} for this propagator in scalar QED,
\bear
\Pi^{(2)}_{\text{scal}} (k^2) &=&
-\frac{e^6}{2(4\pi)^D}
 \int_0^\infty \dfrac{dT}{T^{D+1}} 
e^{-m^2T} 
\int_0^\infty d\bar T
\int_0^Td\tau_a \int_0^Td\tau_b 
 \gamma_{ab}^{D/2}
 \nonumber \\
 && \times
 \int_0^Td\tau_1 \int_0^Td\tau_2
 \, 
\e^{- (G_{12} - \frac{\gamma_{ab}}{4}C^2) k^2}
I\, ,
\ear

\bear
I &=&
D  
(-\ddot G_{ab} + \frac{\gamma_{ab}}{2}\dot G_{ab}^2)
(-\ddot G_{12} - \frac{\gamma_{ab}}{2} \partial_1C\partial_2C)
\frac{1}{k^2}
\nonumber\\ &&
 +
\Bigl[(-\ddot G_{a1} - \frac{\gamma_{ab}}{2}
 \dot G_{ab} \partial_1C)
(-\ddot G_{b2} + \frac{\gamma_{ab}}{2}
\dot G_{ab}  \partial_2C ) + (1\leftrightarrow 2) \Bigr]
\frac{1}{k^2}
\nonumber\\
%&& +
%(-\ddot G_{a2} + \frac{\gamma_{ab}}{2}
%\dot G_{ab} \partial_2 C)
%(-\ddot G_{b1} - \frac{\gamma_{ab}}{2}
%\dot G_{ab} \partial_1 C)
%\nonumber\\
&& - 
[\partial_a C - \frac{\gamma_{ab}}{2} \dot G_{ab} C]
[\partial_b C + \frac{\gamma_{ab}}{2} \dot G_{ab} C]
\Bigl(\ddot G_{12} + \frac{\gamma_{ab}}{2} \partial_1C \partial_2 C \Bigr)
\label{IG}
\ear
($C = G_{1a} - G_{1b} - G_{2a} + G_{2b}$, $ \gamma_{ab} = (\bar T + G_{ab})^{-1}$).

%
%
%
% \begin{figure}
%\centering
%
%\begin{tikzpicture}
%
%\draw (0,0) circle (1);
%\draw (-0.2,0.97) -- (-0.2,-0.97);
%\draw (-0.4, 0.91) -- (-0.4, -0.91);
%\draw (-0.6, 0.8) -- (-0.6, -0.8);
%\node at (0.3,0) {$\dots$};
%\node at (0,-1.5) {$D^{(0)}$};
%
%\end{tikzpicture} \hspace{20pt} \begin{tikzpicture}
%
%\draw (0,0) circle (1);
%\draw (0.2,0.97) -- (0.2,-0.97);
%\draw (-0.7, 0.71) -- (0,-1);
%\draw (0,1) -- (-0.38,0.04);
%\draw (-0.42,-0.06) -- (-0.7,-0.71);
%\node at (0.6,0) {$\dots$};
%\node at (0,-1.5) {$D^{(1)}$};
%
%\end{tikzpicture} \hspace{20pt} \begin{tikzpicture}
%
%\draw (0,0) circle (1);
%\draw (0.2,0.97) -- (0.2,-0.97);
%\draw (-0.7, 0.71) -- (0,-1);
%\draw (0,1) -- (-0.38,0.04);
%\draw (-0.42,-0.06) -- (-0.7,-0.71);
%\draw (-0.4,0.91) -- (-0.4,0.06);
%\draw (-0.4,-0.1) -- (-0.4, -0.91);
%\node at (0.6,0) {$\dots$};
%\node at (0,-1.5) {$D^{(2)}$};
%
%\node at (2.9,0) {$\dots$};
%\node at (2.9,-1.5) {$\cdots$};
%
%\end{tikzpicture}
%
%
%
%
%
%    
%\caption{Caption}
%\label{fig:enter-label}
%\end{figure}
%
%
%

\section{Some remarks on the non-abelian case}

While here we have focussed on the abelian case, let us mention that the integrated-by-parts integrand following from the master formula \eqref{master} 
has recently been found to have other interesting properties that are not visible in the abelian case. In particular, it provides a simple route to the derivation of gluon \cite{140} and graviton \cite{142} Berends-Giele currents.

\end{document}